\catcode`@=11

\newskip\plaincentering \plaincentering=0pt plus 1000pt minus 1000pt
\def\@plainlign{\tabskip=0pt\everycr={}}
\def\eqalignno#1{\displ@y \tabskip\plaincentering
  \halign to\displaywidth{\hfil$\@lign\displaystyle{##}$\tabskip\z@skip
    &$\@lign\displaystyle{{}##}$\hfil\tabskip\plaincentering
    &\llap{$\@lign##$}\tabskip\z@skip\crcr
    #1\crcr}}
\def\leqalignno#1{\displ@y \tabskip\plaincentering
  \halign to\displaywidth{\hfil$\@lign\displaystyle{##}$\tabskip\z@skip
    &$\@lign\displaystyle{{}##}$\hfil\tabskip\plaincentering
    &\kern-\displaywidth\rlap{$\@lign##$}\tabskip\displaywidth\crcr
    #1\crcr}}
\def\plainLet@{\relax\iffalse{\fi\let\\=\cr\iffalse}\fi}
\def\plainvspace@{\def\vspace##1{\noalign{\vskip##1}}}

\def\intic@{\mathchoice{\hskip5\p@}{\hskip4\p@}{\hskip4\p@}{\hskip4\p@}}
\def\negintic@
 {\mathchoice{\hskip-5\p@}{\hskip-4\p@}{\hskip-4\p@}{\hskip-4\p@}}
\def\intkern@{\mathchoice{\!\!\!}{\!\!}{\!\!}{\!\!}}
\def\intdots@{\mathchoice{\cdots}{{\cdotp}\mkern1.5mu
    {\cdotp}\mkern1.5mu{\cdotp}}{{\cdotp}\mkern1mu{\cdotp}\mkern1mu
      {\cdotp}}{{\cdotp}\mkern1mu{\cdotp}\mkern1mu{\cdotp}}}
\newcount\intno@
\def\iint{\intno@=\tw@\futurelet\next\ints@}
\def\iiint{\intno@=\thr@@\futurelet\next\ints@}
\def\iiiint{\intno@=4 \futurelet\next\ints@}
\def\idotsint{\intno@=\z@\futurelet\next\ints@}
\def\ints@{\findlimits@\ints@@}
\newif\iflimtoken@
\newif\iflimits@
\def\findlimits@{\limtoken@false\limits@false\ifx\next\limits
 \limtoken@true\limits@true\else\ifx\next\nolimits\limtoken@true\limits@false
    \fi\fi}
\def\multintlimits@{\intop\ifnum\intno@=\z@\intdots@
  \else\intkern@\fi
    \ifnum\intno@>\tw@\intop\intkern@\fi
     \ifnum\intno@>\thr@@\intop\intkern@\fi\intop}
\def\multint@{\int\ifnum\intno@=\z@\intdots@\else\intkern@\fi
   \ifnum\intno@>\tw@\int\intkern@\fi
    \ifnum\intno@>\thr@@\int\intkern@\fi\int}
\def\ints@@{\iflimtoken@\def\ints@@@{\iflimits@
   \negintic@\mathop{\intic@\multintlimits@}\limits\else
    \multint@\nolimits\fi\eat@}\else
     \def\ints@@@{\multint@\nolimits}\fi\ints@@@}
\def\Sb{_\bgroup\vspace@
        \baselineskip=\fontdimen10 \scriptfont\tw@
        \advance\baselineskip by \fontdimen12 \scriptfont\tw@
        \lineskip=\thr@@\fontdimen8 \scriptfont\thr@@
        \lineskiplimit=\thr@@\fontdimen8 \scriptfont\thr@@
        \Let@\vbox\bgroup\halign\bgroup \hfil$\scriptstyle
            {##}$\hfil\cr}
\def\endSb{\crcr\egroup\egroup\egroup}
\def\Sp{^\bgroup\vspace@
        \baselineskip=\fontdimen10 \scriptfont\tw@
        \advance\baselineskip by \fontdimen12 \scriptfont\tw@
        \lineskip=\thr@@\fontdimen8 \scriptfont\thr@@
        \lineskiplimit=\thr@@\fontdimen8 \scriptfont\thr@@
        \Let@\vbox\bgroup\halign\bgroup \hfil$\scriptstyle
            {##}$\hfil\cr}
\def\endSp{\crcr\egroup\egroup\egroup}
\def\Let@{\relax\iffalse{\fi\let\\=\cr\iffalse}\fi}
\def\vspace@{\def\vspace##1{\noalign{\vskip##1 }}}
\def\aligned{\,\vcenter\bgroup\plainvspace@\plainLet@\openup\jot\m@th\ialign
  \bgroup \strut\hfil$\displaystyle{##}$&$\displaystyle{{}##}$\hfil\crcr}
\def\endaligned{\crcr\egroup\egroup}
\def\matrix{\,\vcenter\bgroup\plainLet@\plainvspace@
    \normalbaselines
  \m@th\ialign\bgroup\hfil$##$\hfil&&\quad\hfil$##$\hfil\crcr
    \mathstrut\crcr\noalign{\kern-\baselineskip}}
\def\endmatrix{\crcr\mathstrut\crcr\noalign{\kern-\baselineskip}\egroup
                \egroup\,}
\newtoks\hashtoks@
\hashtoks@={#}
\def\format{\crcr\egroup\iffalse{\fi\ifnum`}=0 \fi\format@}
\def\format@#1\\{\def\preamble@{#1}%
  \def\c{\hfil$\the\hashtoks@$\hfil}%
  \def\r{\hfil$\the\hashtoks@$}%
  \def\l{$\the\hashtoks@$\hfil}%
  \setbox\z@=\hbox{\xdef\Preamble@{\preamble@}}\ifnum`{=0 \fi\iffalse}\fi
   \ialign\bgroup\span\Preamble@\crcr}

\def\cases{\left\{\,\vcenter\bgroup\plainvspace@
     \normalbaselines\openup\jot\m@th
      \plainLet@\ialign\bgroup$\displaystyle{##}$\hfil&\quad$\displaystyle{{}##}$\hfil\crcr
      \mathstrut\crcr\noalign{\kern-\baselineskip}}

\newif\iftagsleft@
\tagsleft@true
\def\TagsOnRight{\global\tagsleft@false}
\def\tag#1$${\iftagsleft@\leqno\else\eqno\fi
 \hbox{\def\pagebreak{\global\postdisplaypenalty-\@M}%
 \def\nopagebreak{\global\postdisplaypenalty\@M}\rm(#1\unskip)}%
  $$\postdisplaypenalty\z@\ignorespaces}
\interdisplaylinepenalty=\@M
\def\plainallowdisplaybreak@{\def\allowdisplaybreak{\noalign{\allowbreak}}}
\def\plaindisplaybreak@{\def\displaybreak{\noalign{\break}}}
\def\align#1\endalign{\def\tag{&}\plainvspace@\plainallowdisplaybreak@\plaindisplaybreak@
  \iftagsleft@\plainlalign@#1\endalign\else
   \plainralign@#1\endalign\fi}
\def\plainralign@#1\endalign{\displ@y\plainLet@\tabskip\plaincentering\halign to\displaywidth
     {\hfil$\displaystyle{##}$\tabskip=\z@&$\displaystyle{{}##}$\hfil
       \tabskip=\plaincentering&\llap{\hbox{\rm(##\unskip)}}\tabskip\z@\crcr
             #1\crcr}}
\def\plainlalign@
 #1\endalign{\displ@y\plainLet@\tabskip\plaincentering\halign to \displaywidth
   {\hfil$\displaystyle{##}$\tabskip=\z@&$\displaystyle{{}##}$\hfil
   \tabskip=\plaincentering&\kern-\displaywidth
        \rlap{\hbox{\rm(##\unskip)}}\tabskip=\displaywidth\crcr
               #1\crcr}}

\def\re@#1{\par\hangindent\parindent\indent\llap{#1\enspace}\ignorespaces}
\def\qfootnote#1{\edef\@sf{\spacefactor\the\spacefactor}{}#1\@sf
      \insert\footins{\let\egroup=}\footnotesize 
      \interlinepenalty100 \let\par=\endgraf
        \leftskip=0pt \rightskip=0pt
        \splittopskip=10pt plus 1pt minus 1pt \floatingpenalty=20000
   \smallskip\re@{#1}\bgroup\strut\aftergroup{\strut\egroup}\let\next}
\catcode`\@=\active

\documentstyle[11pt]{article}
\begin{document}

\newcommand{\be}{\begin{equation}}
\newcommand{\ee}{\end{equation}}

\newcommand{\ba}{\begin{equation} \aligned}
\newcommand{\ea}{\endaligned \end{equation}}

\normalsize \baselineskip 18pt

\title{\bf Spinor Field Realizations of Non-critical $W_{2,s}$ Strings
           \thanks{Project supported by the National Natural Science Foundation of China(No. 10275030).
            }}
\author{ Y. S. Duan\thanks{E-mail address: ysduan@lzu.edu.cn} $\;\;$
         Y. X. Liu \thanks{Corresponding author; E-mail address: liuyx01@st.lzu.edu.cn} $\;\;$
         L. J. Zhang\thanks{E-mail address: zhanglj02@st.lzu.edu.cn} $\;\;$
        \\
Institute of Theoretical Physics, Lanzhou University, Lanzhou
730000, P. R. China}
\date{}
\maketitle

\begin{center}
\begin{minipage}{120mm}
\vskip 0.5in

{\bf\noindent Abstract} \\

\indent In this paper, we construct the nilpotent
Becchi-Rouet-Stora-Tyutin($BRST$) charges of spinor non-critical
$W_{2,s}$ strings. The cases of $s=3,4$ are discussed in detail,
and spinor realization for $s=4$ is given explicitly. The $BRST$
charges are graded. \\


\end{minipage}
\end{center}

\newpage
{\bf\noindent  1. Introduction} \\
\indent As is well known, $W$ algebra has found remarkable
applications in $W$ gravity and $W$ string theories since its
discovery in 1980's [1,2]. Furthermore, it appears in the quantum
Hall effect, black holes, in lattice models of statistical
mechanics at criticality, and in other physical models [3,4]
and so on.\\
\indent The $BRST$ formalism [5] has turned out to be rather
fruitful in the study of string theories and the study of critical
and non-critical $W$ string theories. The $BRST$ charge for
$W_{3}$ string was first constructed in [6],
 and the detailed studies of it can be found in [6-12]. A natural generalization of the
$W_{3}$ string, i.e. the critical $W_{2,s}$ string, is a
higher-spin string with local spin-2 and spin-$s$ symmetries on
the world-sheet. Critical $BRST$ charges for such theories have
been constructed for $s=4,5,6,7$ [13-15]. For a system with
non-linear symmetry, however, the critical and non-critical $BRST$
charges are intrinsically different. In [16,17], a $BRST$ charge
for the so-called non-critical $W_{3}$ string was found. Such
$W_{3}$ string is a theory of $W_{3}$ matter coupled to $W_{3}$
gravity; it is a generalization of two-dimensional matter couple
to gravity. Shortly, a non-critical $BRST$ charge for the
$W_{2,4}$ string was constructed by generalizing the $W_{3}$
results of [16-18]. And some of the physical states, including
some of the ghost-number
zero ground-ring generators, were obtained.\\
\indent However, all of these theories about $W_{2,s}$ strings
mentioned above are based on scalar field. In the work [19], we
pointed out the reason that the scalar $BRST$ charge is difficult
to be generalized to a general $W_{N}$ string. At the same time,
we found the methods to construct the spinor field realization of
critical $W_{2,s}$ strings and $W_{N}$ strings. Assuming the
$BRST$ charges of the $W_{2,s}$ strings and $W_{N}$ strings are
graded, we studied the exact spinor field realizations of
$W_{2,s}(s=3,4,5,6)$ strings and $W_{N}(N=4,5,6)$ strings [19-22]
by using our program.\\
\indent Since so far there is no work focussed on the research of
spinor non-critical $W_{2,s}$ strings, we will construct the
nilpotent $BRST$ charges of spinor non-critical $W_{2,s}$ strings
by taking into account the property of spinor field in this paper.
To construct a $BRST$ charge one must first solve the forms of
$T_{M}$ and $W_{M}$ determined by the OPEs of $TT$, $TW$ and $WW$.
Then direct substitution of these results into $BRST$ charge leads
to the grading realizations. Such constructions are discussed for
$s=3,4$. These results will be of importance for constructing
super $W$ strings, and they provide the
essential ingredients.\\
\indent This paper is organized as follows. We begin in Sec. 2 by
reviewing spinor field realizations of critical $W_{2,s}$ strings
which were obtained in [19-22]. Then we give the grading $BRST$
method to construct spinor field $BRST$ charges of non-critical
$W_{2,s}$ strings. Subsequently, we mainly discuss the spinor
field realizations of the non-critical $W_{2,3}$ string and
$W_{2,4}$ string, and construct a non-critical $BRST$ charge for
the $W_{2,4}$ string. And finally,
a brief conclusion is given.\\

{\bf\noindent  2. Review of spinor field realizations of critical $W_{2,s}$ strings} \\
\indent In [20] the spinor $BRST$ charges were constructed for
critical $W_{2,s}$ strings theories, that is, for pure
$W_{2,s}$-matters. The authors introduce the $(b,c)$ ghost system
for the spin-2 current, and the $(\beta,\gamma)$ for the spin-$s$
current, where $b$ has spin 2 and $c$ has spin -1 whilst $\beta$
has spin $s$ and $\gamma$ has spin $(1-s)$. The ghost fields
$b,c,\beta,\gamma$ are all
bosonic and communicating. They satisfy the OPEs\\
\be b(z)c(\omega)\sim \frac{1}{z-\omega},\;\;\;
\beta(z)\gamma(\omega)\sim \frac{1}{z-\omega}, \ee \noindent in
the other case the OPEs vanish. The spinor field $\psi$ has spin
1/2 and satisfies the OPE
 \be \psi(z)\psi(\omega)\sim
-\frac{1}{z-\omega}. \ee \noindent Then the $BRST$ charge for the
spin-2 plus spin-$s$ string takes the form: \be
    Q_{B}=Q_{0}+Q_{1},
\ee \be
    Q_{0}=\oint dz\; c(T^{eff}+T_{\psi}+KT_{bc}+yT_{\beta\gamma}),
\ee \be
    Q_{1}=\oint dz\; \gamma F(\psi,\beta,\gamma),
\ee \noindent where $K,y$ are pending constants and the operator
$F(\psi,\beta,\gamma)$ has spin $s$ and ghost number zero. The
energy-momentum tensors in (4) are given by \be
    T_{\psi}=-\frac{1}{2}\partial\psi\psi,
\ee \be
    T_{\beta\gamma}=s\beta\partial\gamma+(s-1)\partial\beta\gamma,
\ee \be
    T_{bc}=2b\partial c+\partial bc,
\ee \be
    T^{eff}=-\frac{1}{2}\eta_{\mu\nu}\partial Y^{\mu} Y^{\nu}.
\ee \noindent  The $BRST$ charge is graded with
$Q_{0}^{2}=Q_{1}^{2}=\{Q_{0},Q_{1}\}=0$. The first condition
$Q_{0}^{2}=0$ is satisfied for an arbitrary $s$. The remaining two
nilpotency conditions determine the precise form of the operator
$F(\psi,\beta,\gamma)$ and the exact $y$. They have no constraint
for the coefficient $K$. The particular method used to construct
$F(\psi,\beta,\gamma)$ can be found in [20,22]. We have obtained
the solutions for $s=3,4,5,6$ and
discussed the case of any $s$ [19-22].\\

{\bf\noindent  3. Spinor field realizations of non-critical $W_{2,s}$ strings} \\
\indent The non-critical $W_{2,s}$ strings are the theories of
$W_{2,s}$ gravity coupled to a matter system on which the
$W_{2,s}$ algebras are realized. Now we give the spinor field
realizations of them.\\
\indent The $BRST$ charge takes the form: \be
    Q_{B}=Q_{0}+Q_{1},
\ee \be
    Q_{0}=\oint dz\; c(T^{eff}+T_{\psi}+T_{M}+KT_{bc}+yT_{\beta\gamma}),
\ee \be
    Q_{1}=\oint dz\; \gamma F(\psi,\beta,\gamma,T_{M},W_{M}),
\ee

\noindent where the matter currents $T_{M}$ and $W_{M}$, which
have spin 2 and $s$ respectively, generate the $W_{2,s}$ algebra,
whilst the energy-momentum tensors
$T_{\psi},T_{\beta\gamma},T_{bc}$ and $T^{eff}$ are given by
(6-9). The $BRST$ charge generalizes the one for the scalar
non-critical $W_{2,s}$ strings, and it is also graded with
$Q_{0}^{2}=Q_{1}^{2}=\{Q_{0},Q_{1}\}=0$. Again the first condition
is satisfied for any $s$, and the remaining two conditions
determine the coefficients of the terms in
$F(\psi,\beta,\gamma,T_{M},W_{M})$ and $y$.\\
\indent In order to build the spinor non-critical $W_{2,s}$
strings theories, we need the explicit realizations for the matter
currents $T_{M}$ and $W_{M}$. Generally, the two-spinor
realizations of the $W_{2,s}$ algebras take the form:
\ba T_{M} &
= T_{Y_{1}}+T_{Y_{2}}=-\frac{1}{2}\partial Y_{1}
Y_{1}-\frac{1}{2}\partial Y_{2} Y_{2}, \\
W_{M} & = W_{Y_{1}}+W_{Y_{1}}+W_{Y_{1},Y_{2}}, \ea \noindent where
$W_{Y_{i}}(i=1,2)$ indicates the terms which are formed by
$Y_{i}$, and $W_{Y_{1},Y_{2}}$ corresponds to that of $Y_{1}$ and
$Y_{2}$. In conformal OPE language the $W_{2,s}$ algebra takes the
following forms:\\
\ba T(z)T(\omega) \sim &  \frac{C/2}{(z -
w)^4}+\frac{2T(\omega)}{(z-\omega)^2}+ \frac{\partial T(\omega)}{(z-\omega)}, \\
T(z)W(\omega) \sim &  \frac{s W(\omega)}{(z-\omega)^2}+
\frac{\partial
W(\omega)}{(z-\omega)},\\
W(z)W(\omega) \sim & \frac{C/s}{(z - w)^{2s}} + \sum _{\alpha}
\frac{P_{\alpha}(\omega)}{(z-\omega)^{\alpha+1}}, \ea \noindent in
which $P_{\alpha}(\omega)$ are polynomials in the primary fields
$W$, $T$ and their derivatives. For an exact $s$, the precise form
of $W$ and the corresponding central charge $C$ can be solved by
means of these OPEs. Substituting them into (12), we can get the
final result of $Q_{1}$. So, the $BRST$ charge $Q_{B}$ for
the spinor non-critical $W_{2,s}$ string can be obtained.\\

{\bf\noindent  4. The particular results} \\
\indent In present section, using the grading $BRST$ method and
the procedure mentioned in Sec. 3, we will discuss the exact
solutions of spinor field realizations of the non-critical
$W_{2,s}$ strings
for the cases of s=3 and 4.\\

{\noindent 4.1. Spinor field realization of the non-critical $W_{2,3}$ string}\\
\indent In this case, $Q_{B}$ takes the form of (10). The most
extensive combinations of $F$ in (12) with correct spin and ghost
number can be constructed as following: \ba
F(\psi,\beta,\gamma,T_{M},W_{M}) & =
f3[1]\beta^3\gamma^3+f3[2]\partial\beta\beta\gamma^2
                     + f3[3]\partial \beta \partial \gamma
                    +f3[4] \beta\gamma\partial\psi\psi + f3[5] \beta
                    \partial ^2\gamma \\
           & +f3[6]\partial ^2\psi\psi  + f3[7]\beta\gamma T_{M}  + f3[8]\partial\psi\psi T_{M} + f3[9]\partial T_{M} + f3[10]W_{M}. \\
\ea \noindent Substituting (15) back into (12) and imposing the
nilpotency conditions, we can determine $y$ and
$f3[i](i=1,2,...10)$. They correspond to three sets of solutions,
respectively. i.e.\\
\noindent (i) $y=0$ and
$$\align
  f3[4] &=  f3[6] =  f3[7] = f3[8] = f3[9] = f3[10] = 0,
\endalign $$
\noindent and $f3[1],f3[2],f3[3],f3[5]$ are arbitrary constants
but do not vanish at the same
time.\\
\noindent (ii) $y=1$ and
$$\align
  f3[1] & = \frac{1}{150}(-7 f3[3] + 3 f3[5]),\;\;\;  f3[2] = \frac{1}{15}(7 f3[3] - 3
  f3[5]),\\\allowdisplaybreak
  f3[4] &= \frac{22}{5}f3[3] - \frac{78}{5} f3[5] + Cm f3[7],\;\;\;  f3[6] = -11 f3[3] + 39 f3[5] - \frac{5}{2} Cm f3[7],\\\allowdisplaybreak
  f3[8] &= 0,\;\;\; f3[9] = -\frac{5}{2}f3[7],\\
\allowdisplaybreak
\endalign $$
\noindent where $f3[3],f3[5],f3[7]$ and $f3[10]$ are arbitrary
constants but do not vanish at the same time. $Cm$ is the matter central charge corresponding to the matter currents $T_{M}$ and $W_{M}$.\\
\noindent (iii) $y$ is an arbitrary constant and
$$\align
f3[4] & =  f3[6] =  f3[7] = f3[8] = f3[9] = f3[10] =
0,\\\allowdisplaybreak
 f3[1] &= -\frac{8}{55}f3[5],\;\;\; f3[2] = \frac{16}{11} f3[5],\;\;\;  f3[3] = \frac{39}{11}
 f3[5],\\\allowdisplaybreak
\endalign $$
\noindent where $f3[5]$ is an arbitrary constant but does not vanish.\\
\indent Now we turn to the construction of the form of $W$. The
OPE $W(z)W(\omega)$ in (14) is given by [1]\\
 \ba W(z)W(\omega) \sim & \frac{C/3}{(z -
w)^6} +
\frac{2T}{(z-\omega)^4}+ \frac{\partial T}{(z-\omega)^3}\\
+& \frac{1}{(z-\omega)^2}(2 \Theta \Lambda +\frac{3}{10}\partial
^2 T )+\frac{1}{(z-\omega)}(\Theta
\partial \Lambda +\frac{1}{15}\partial ^3 T),
\ea
where\\
 \ba \Theta= \frac{16}{22+5C},\;\;\;\;\;\; \Lambda = T^2
- \frac{3}{10}
\partial ^2 T.
\ea
\noindent We can write down the most general possible
structure of $W_{M}$ for $s=3$ :
 \ba W_{M} & = g3[1]
\partial^2 Y_{1} Y_{1} + g3[2]
\partial^2 Y_{2} Y_{2}+ g3[3] \partial^2 Y_{1} Y_{2} + g3[4] Y_{1}
\partial^2 Y_{2}+ g3[5] \partial Y_{1} \partial Y_{2}.\ea
\noindent Unfortunately, we find there is no nontrivial solution
for the two-spinor realization of $W_{2,3}$ algebra.\\

{\noindent 4.2. Spinor field realization of the non-critical $W_{2,4}$ string}\\
\indent Similarly, for the case $s=4$, $Q_{B}$ also takes the form
of (10) and $F$ can be expressed in the following form:
\ba    F(\psi,\beta,\gamma,T_{M},W_{M}) & = f4[1]\beta ^4\gamma ^4+f4[2](\partial \beta)^2\gamma ^2+f4[3] \beta ^3\gamma ^2\partial \gamma + \ f4[4]\beta^2 (\partial \gamma )^2   \\
                & +f4[5] \beta ^2\gamma ^2 \partial \psi \psi +f4[6]\partial \beta \gamma \partial \psi \psi+ f4[7]\beta \partial \gamma \partial \psi \psi +f4[8] \beta  \partial ^2\beta  \gamma ^2 \\
                & +f4[9] \partial ^2\beta \partial \gamma + \ f4[10]\partial \beta \partial ^2\gamma +f4[11] \partial ^2\psi \partial \psi +f4[12] \beta \partial ^3\gamma    \\
                & + f4[13] \partial ^3\psi \psi + f4[14]\partial \beta \gamma T_{M} + f4[15] \beta \partial \gamma T_{M} + f4[16]\partial\psi\psi T_{M}  \\
                & + f4[17]\beta\gamma\partial T_{M} + f4[18]T_{M}^2 + f4[19]\partial^2 T_{M} +f4[20]W_{M} . \\
\ea \indent There are three sets of solutions:\\
\noindent (i) $y=0$ and
$$\align
f4[5] &= f4[6] = f4[7] = f4[11] = f4[13] = f4[16] = f4[18] =
f4[19] = f4[20] = 0,\\ \allowdisplaybreak
f4[14] &= f4[17] = \frac{1}{2}f4[15],\\
\allowdisplaybreak
\endalign $$
\noindent where
$f4[1],f4[2],f4[3],f4[4],f4[8],f4[9],f4[10],f4[12],f4[15]$ are
arbitrary constants but do not vanish at the same
time.\\
\noindent (ii) $y=1$ and
$$\align
f4[1] & = \frac{1}{468930} (58 f4[6] + 3 (38 f4[7] +  21 (289
f4[8] + 5 Cm (-2 f4[14] + f4[15])))),\\ \allowdisplaybreak
f4[2]&=\frac{1}{15312}(116f4[6]+63 f4[7] + 6 (2571 f4[8] + 50 Cm
(-2f4[14] + f4[15]))),\\\allowdisplaybreak f4[3] & =
\frac{1}{20097} (58 f4[6] +
3 (38 f4[7] + 21 (289 f4[8] + 5 Cm (-2 f4[14] +   f4[15])))),\\
\allowdisplaybreak
f4[4] & =  \frac{1}{232}(5 f4[7] + 2 (57 f4[8] +  5 Cm (-2 f4[14] + f4[15]))),\\
\allowdisplaybreak
f4[5] & = \frac{1}{21}(-4 f4[6] + 3 f4[7]),\\
\allowdisplaybreak f4[9] & = \frac{1}{696} (7 f4[7] + 1134 f4[8] +
232 f4[10] - 28 Cm f4[14] + 14 Cm f4[15]),\\ \allowdisplaybreak
f4[11] & = \frac{1}{70} (330 f4[6] + 10 f4[7] +  3 Cm (278 f4[14]
- 139 f4[15] + (22 + 5 Cm) f4[18])),\\ \allowdisplaybreak f4[12] &
= \frac{1}{1392} (-35 f4[7] + 2
(-399 f4[8] + 464 f4[10] + 70 Cm f4[14] -  35 Cm f4[15])),\\
\allowdisplaybreak f4[13] & =  \frac{1}{210} (-110 f4[6] + 160
f4[7] - Cm
(278 f4[14] -  139 f4[15] + (22 + 5 Cm) f4[18])),\\
\allowdisplaybreak
f4[16] & =  \frac{2}{5}(278 f4[14] - 139 f4[15] + (22 + 5 Cm) f4[18]),\\
\allowdisplaybreak
f4[17] & = -2 f4[14] + \frac{3}{2}f4[15],\\
\allowdisplaybreak
f4[19] & = \frac{1}{10}(28 f4[14] - 14 f4[15] - 3 f4[18]),\\
\allowdisplaybreak
\endalign $$
\noindent where
$f4[6],f4[7],f4[8],f4[10],f4[14],f4[15],f4[18],f4[20]$ are
arbitrary constants but do not vanish at the same
time.\\
\noindent (iii) $y$ is an arbitrary constant and
$$\align
f4[5] &= f4[6] = f4[7] = f4[11] = f4[13] = f4[16] = f4[18] =
f4[19] = f4[20] = 0,
\\ \allowdisplaybreak
f4[1] &= \frac{867}{22330}f4[8],\;\; f4[2] =
\frac{2571}{2552}f4[8], \;\;  f4[3] = \frac{289}{319}f4[8],
\;\;  f4[4] = \frac{57}{116}f4[8], \\
\allowdisplaybreak f4[9] &= \frac{189}{116}f4[8] +
\frac{1}{3}f4[10],
\;\;\;  f4[12] = -\frac{133}{232}f4[8] + \frac{2}{3}f4[10], \\
\allowdisplaybreak f4[14] &= f4[17] = \frac{1}{2} f4[15], \\
\allowdisplaybreak
\endalign $$
\noindent where $f4[8],f4[10],f4[15]$ are arbitrary constants but
do not vanish at the same
time.\\
\indent Note that the coefficient $f4[20]$ for the term of $W_{M}$
in $F(\psi,\beta,\gamma,T_{M},W_{M})$ is zero for the first and
last solutions, we only need the result with
$y=1$ for non-critical $W_{2,4}$ string.\\
\indent Let us now consider the construction of $W$ with $s=4$.
The OPE $W(z)W(\omega)$ for $W_{2,4}$ algebra is given by [23,24]:
\ba W(z)W(\omega) \sim &\{\frac{2 T}{(z - w)^6}+\frac{\partial
T}{(z-\omega)^5}+\frac{3}{10} \frac{\partial ^2
T}{(z-\omega)^4}\\
+ & \frac{1}{15}\frac{\partial ^3
T}{(z-\omega)^3}+\frac{1}{84}\frac{\partial ^4
T}{(z-\omega)^2}+\frac{1}{560}\frac{\partial ^5
T}{(z-\omega)}\}\\
+ & b_{1} \{\frac{U}{(z-\omega)^4}+\frac{1}{2} \frac{\partial
U}{(z-\omega)^3}+ \frac{5}{36} \frac{\partial ^2
U}{(z-\omega)^2}+\frac{1}{36} \frac{\partial ^3
U}{(z-\omega)}\}\\
+ & b_{2} \{ \frac{W}{(z-\omega)^4}+\frac{1}{2} \frac{\partial
W}{(z-\omega)^3}+ \frac{5}{36} \frac{\partial ^2
W}{(z-\omega)^2}+\frac{1}{36} \frac{\partial ^3 W}{(z-\omega)}\}\\
+& b_{3} \{ \frac{G}{(z-\omega)^2}+\frac{1}{2}\frac{\partial
G}{(z-\omega)} \}+ b_{4}\{ \frac{A}{(z-\omega)^2}+\frac{1}{2}
\frac{\partial A}{(z-\omega)}\}\\
+ & b_{5}\{ \frac{B}{(z-\omega)^2}+\frac{1}{2} \frac{\partial
B}{(z-\omega)}\}+\frac{C/4}{(z-\omega)^8}, \ea \noindent where the
composites U (spin 4), and G, A and B (all spin 6 ), are defined
by \ba U & =(TT)-\frac{3}{10} \partial ^2 T,\;\;\;\;\;\;
G=(\partial ^2 T T)-\partial (\partial T T)+\frac{2}{9} \partial
^2
(TT)-\frac{1}{42}\partial ^4 T,\\
A & =(TU)-\frac{1}{6}\partial ^2 U, \;\;\;\;\;\;
B=(TW)-\frac{1}{6}\partial ^2 W ,\ea
\noindent with normal
ordering of products of currents understood. The coefficients
$b_{1},b_{2},b_{3},b_{4}$ and $b_{5}$ are given by
\ba
 b_{1} & =\frac{42}{5C+22},\;\;\;\;\;\;
b_{2}=\sqrt{\frac{54(C+24)(C^2 -172C
+196)}{(5C+22)(7C+68)(2C-1)}},\\
b_{3} & =\frac{3(19C-524)}{10(7C+68)(2C-1)},\;\;\;\;\;\; b_{4}
=\frac{24(72C+13)}{(5C+22)(7C+68)(2C-1)},\\
b_{5} & = \frac{28}{3(C+24)} b_{2}. \ea \noindent These relations
determine the coefficients of the terms in $W_{M}$, the result
turns out to be very simple as follows: \ba C_{M}=1,
 \ea \ba W_{M}
& =  \frac{1}{72\sqrt{6}}(\partial^3 Y_{1} Y_{1} -9
\partial^2 Y_{1} \partial Y_{1} + 84 \partial Y_{1} Y_{1} \partial Y_{2} Y_{2} +
\partial^3 Y_{2} Y_{2} -9 \partial^2 Y_{2} \partial Y_{2}).
\ea
\noindent Substituting these results into (12) gives the final
spinor field realization of the non-critical $W_{2,4}$ string.\\

{\bf\noindent 5. Conclusion}\\
\indent In this paper, the spinor field realizations of
non-critical $W_{2,s}$ strings have been studied. We have
discussed the cases of $s=3,4$ in detail and constructed the
$BRST$ charges for $s=4$ by explicit computation. The construction
was based on demanding nilpotence of the $BRST$ charges, making no
reference to whether or not an underlying $W_{2,s}$ algebra
exists. These solutions are very standard, that is, there are
three solutions for $s=3$ and 4, respectively. We find that the
OPE of the spin-3 current with itself gives rise only to a null
current, so the two-spinor non-critical realization of $W_{2,3}$
string is believed not to exist. For the case $s=4$, $W_{M}$ has
one simple solution. Of course more spinor realizations of
$W_{2,s}$ strings can be calculated with our procedure. We expect
that there should exist such realizations with higher spin $s$.
Having obtained the non-critical abstract $BRST$ charges for
$W_{2,s}$ strings, we can investigate the
implications for the corresponding string theories.\\

{\bf\noindent Acknowledgements}\\
\indent It is a pleasure to thank Dr. H.Wei for discussion. We
have also made extensive use of a Mathematica package for
calculating OPEs, written by Prof. K. Thielemans [25].

\end{document}